\documentclass[twocolumn,pre,aps,preprintnumbers,showkeys,amsmath,amssymb]{revtex4-2}

\usepackage{graphicx}
\usepackage{amsmath}
\usepackage{inputenc}
\usepackage{mathtools}
\usepackage{amssymb}
\usepackage{braket}
\usepackage{setspace}
\usepackage{makeidx}
\usepackage{bm}
\usepackage{color}
\usepackage{hyperref}
\usepackage{natbib}
\usepackage{placeins}

\begin{document}
	
	\title{Non-Markovian effects on the steady state properties of a damped harmonic oscillator}
	
	\author{Faisal Farooq}
	\author{Irfan Ahmad Dar}
	\author{Muzaffar Qadir Lone}
	\thanks{Corresponding author: lone.muzaffar@uok.edu.in}
	
	\affiliation{Quantum Dynamics Lab, Department of Physics, University of Kashmir, Srinagar-190006 India}
	\begin{abstract}
		We analyze the steady-state characteristics of a damped harmonic oscillator (system) in presence of a non-Markovian bath characterized by Lorentzian spectral density. Although Markovian baths presume memoryless dynamics, the introduction of complex temporal connections by a non-Markovian environment radically modifies the dynamics of the system and its steady-state behaviour. We obtain the steady-state Green's functions and correlation functions of the system using the Schwinger-Keldysh formalism. In both rotating and non-rotating wave approximation, we analyzed various emergent properties like effective temperature and distribution function. We also explore the impact of dissipation and non-Markovian bath on the quantum Zeno and anti-Zeno effects. We show that a transition between Zeno to anti-Zeno effect can be tuned by bath spectral width and the strength of dissipation.  
	\end{abstract}	
	\maketitle	
	\section{Introduction}
	
	 The field of open quantum systems\cite{weiss} present the most fascinating challenges, not only from a fundamental perspective but also due to their relevance in practical applications and experimental implementations \cite{von,hp,wt,rivas}. This class of problems provides a captivating interaction of coherent Hamiltonian dynamics with the incoherent dissipative dynamics resulting from  the bath \cite{diehlqsp,muller}. From a fundamental perspective, dissipative processes are now viewed as a resource that enables engineering of quantum states of matter \cite{diehlqsp,muller,cirac} and the facilitation of quantum transport features \cite{francois,QT1}, after previously being thought of as a nuisance since they destroy coherence \cite{krauss,roland,walls}. Particle losses have been a particularly important kind of dissipative process, and recent study in cold atomic systems has made them extremely controllable. For instance, this is the situation when local losses are realized in fermionic systems\cite{lebrat} or weakly interacting Bose gases\cite{barontini,corman,huang}.
	 
	 The damped harmonic oscillator is a fundamental model for studying dissipative processes in quantum systems \cite{caldeira,caldeira2,ir}. It describes the interplay between a system's inherent oscillatory motion and its interaction with an environment, often represented by a bath of harmonic oscillators \cite{weiss}. This coupling leads to energy dissipation and decoherence, which are central to understanding open quantum systems. The dynamics of a damped oscillator provides an insight into phenomena such as relaxation, correlation functions, and steady-state behaviors \cite{hp,ingold}. Coupling of a harmonic oscillator to a non-Markovian environment \cite{maghrebi} gives rise to complex behavior, including intricate interactions between the system and the environment, and the emergence of non-classical effects \cite{samyadeb}. This is due to memory effects and information backflow from the environment to the system \cite{rivas2}, which leads to non-exponential relaxation of the quantum coherence \cite{hp2} and  modified steady-state properties  \cite{ask}. All these are essential to understand and control  in the applications of   quantum systems, which include quantum information processing \cite{nc} and quantum thermodynamics \cite{whitney}. 
	 
	  There has been a growing interest in studying the Markovian and non-Markovian dynamics in open quantum systems due its important applications in quantum information science\cite{hp2,rivas2,mehboob,rayees}.  Non-Markovian effects arise due to back flow of information from bath to the system while Markovian effects see no retrieval of information from the bath\cite{hp,rivas2}. In this work, we consider a quantum system represented by a bosonic mode subject to the decay via Lindblad dynamics. We call this system a damped oscillator.  Such a system is a variant of Dicke model in the large $N$-limit\cite{nagy,nagy2}. Furthermore, we couple the system to a non-Markovian bath with Lorentzian spectral density. We consider both non-rotating and rotating wave approximations of the model and analyze various  properties in the steady state of the system like distribution function, effective temperature, quantum Zeno (QZ) and anti-Zeno (AQZ) effects\cite{gagen,helmer}. QZ effect describe a situation where the dynamics of a given system can be  frozen with continuous measurements while the accelerating the induced dynamics representing AQZ effect\cite{khalfin,missud}.
	 Therefore, it would be interesting to look for non-Markovian effects on the steady properties of the system and find the transition from QZ to AQZ by tuning non-Markovinity in the system.

	The theoretical study of dissipative phenomena poses a significant challenge, and various methods have been employed to address the interaction of the system with its environment (bath) \cite{weiss,von,hp,wt,rivas,diehlqsp}.  A commonly used approach involves utilizing non-Hermitian Hamiltonians \cite{yamamoto,nakagawa,yamamoto2}. These non-Hermitian models can be subjected to a variety of field-theoretic and numerical techniques \cite{yamamoto,nakagawa}, such as bosonization  and the renormalization group analysis \cite{yamamoto2,yamamoto3}. However, it is challenging to determine how accurately these treatments can capture the dynamics of the system observables because they frequently ignore or, at most, address the  quantum-jump element of the Lindblad master equation in an approximate manner. Other methods assume that the bath is Markovian and address the entire Lindblad master equation \cite{francois,damanet2,mazza}. In this work, we focus on the Schwinger-Keldysh (SK) functional technique to tackle the  non-equilibrium dynamics \cite{kamenev,torre,sieberer}. Such approach has found its applications in condensed matter systems \cite{dunnet,yongxin,irfan}, cosmology \cite{Barvinsky,barvinsky2} and string theory \cite{jan,matteo}.

	This paper is organized in the following way. Section II introduces model calculations in non-rotating wave approximation followed by section III on quantum Zeno effect. Section IV deals  rotating wave approximation. Finally, we conclude in section V.

\section{Model Calculations}\label{model calculations}
The dynamics of a dissipative quantum system is usually given by Lindblad equation ($\hbar=1$):
\begin{eqnarray}
	\frac{d \rho}{dt}=-{\rm i}[H,\rho] + \mathcal{L}(\rho).
\end{eqnarray}
In this equation, $\rho$ represents the density matrix for the system under consideration. The term $-{\rm i}[H,\rho]$ represents the coherent evolution of the system while the term $  \mathcal{L}(\rho)$ describes dissipation in the system due to various processes. For the Lindblad dynamics it has the form 
\begin{eqnarray}
	\mathcal{L}(\rho)= \sum_i [2 L_i \rho L_i^{\dagger}-\{L_i^{\dagger}L_i,\rho\}],
\end{eqnarray}
with $L_i$ as the Lindblad operator. In our case, we assume Lindblad operator $L =\sqrt{\Gamma_M} a$  ($a$ is the annihilation operator of the system) corresponding to particle loss at rate $\Gamma_M$, so that we can write   
 $\mathcal{L}(\rho)= \Gamma_M  \ [2 a \rho a^{\dagger}  -\{a^{\dagger} a,\rho\}] $.
We now consider our system, damped harmonic oscillator modeled as a bosonic mode subject to Lindblad dynamics with $L =\sqrt{\Gamma_M} a$. Also, we couple this system to a  non-Markovian bath. Such a system is a variant of Dicke Model. The total Hamiltonian in the non-rotating wave approximation  can be written as follows
\begin{eqnarray}
	H &=& H_A + H_B + H_{AB}    \nonumber  \\ 
	&=&  \omega_0 a^{\dagger} a
	+ \sum_k \omega_k b_k^{\dagger} b_k+ \sum_{k}(a+a^{\dagger})(g_k b_k + g_k^* b_k^{\dagger}),\nonumber\\
	\label{model}
\end{eqnarray}
$H_{A} $ describes the  free Hamiltonian of the system with annihilation (creation) operators as $a$ ($ a^{\dagger} $) and $ \omega_0 $ as the characteristic frequency of the bosonic mode. The second term $H_B$ describes the bath with annihilation (creation) operators $ b_k $ ($ b_k^{\dagger} $) for the $ k $-th bath mode and energy $ \omega_k $. The final term $ H_{AB}$ captures the interaction of the system with bath and  $ g_k $ denotes the coupling strength of $k$-th bath mode and system. In the rotating wave approximation, the terms $ab_k$ and $a^{\dagger}b_k^{\dagger}$ are ignored.   This model can be obtained from the large $N$ limit of the Dicke model, where an ensemble of two level atoms are interacting with a bath \cite{nagy,nagy2,hepp}. The system part $H_A$ therefore represents the collective bosonic mode as a result of thermodynamic limit. This model has been realized in various settings like dissipative Bose-Einstein condensates \cite{konya}, multilevel atom-schemes\cite{dimer,ritsch}, etc.

\subsection{Keldysh Field Theory}\label{KFT}
In this section, we use SK field theoretic technique to investigate the dynamics and calculate various observables in our model. The idea of SK functional technique is to evaluate the time evolution on a closed time contour with a forward and backward branch such that  value of the fields match at time $t=\infty$. Let $|\varPhi\rangle$ be the coherent state representing some general quantum system, then the SK action is defined through the partition function \cite{sieberer,torre,kamenev}
{
\begin{eqnarray}
	Z=\int{\rm D} [\varPhi^*_+,\varPhi_+,\varPhi^*_-,\varPhi_-] {\rm e}^{{\rm i}S_{SK}[\varPhi^*_+,\varPhi_+, \varPhi^*_-, \varPhi_-]},
\end{eqnarray}
where the integration measure is given by 
\begin{eqnarray}
  {\rm D} [\varPhi^*_+,\varPhi_+,\varPhi^*_-,\varPhi_-]=\lim_{N \to \infty}
\prod_{n=0}^N \frac{{\rm d}\varPhi^*_+ {\rm d}\varPhi_+}{\pi} \frac{{\rm d}\varPhi^*_-{\rm d}\varPhi_-}{\pi}.
\end{eqnarray}}
$S_{SK}$ represents the SK action and the fields on forward and backward branch of the Keldysh contour are given by $\varPhi_+, \varPhi_-$ respectively. $\varPhi^*$ is the complex conjugate of $\varPhi$. This action can be further simplified using the Keldysh rotation by defining new fields $\varPhi_{c}=\frac{\varPhi_+ +\varPhi_- }{\sqrt{2}}$ and  $\varPhi_{q}=\frac{\varPhi_+ -\varPhi_- }{\sqrt{2}}$. $\varPhi_c, \varPhi_q$ are known as classical and quantum fields respectively. This is attributed to the fact that $\varPhi_c$ has a non-vanishing expectation value while $\varPhi_q$ does not have. The same is true for conjugate fields. Next, we simplify our problem using the coherent states for the system and bath. Let  $|\zeta\rangle$ be the coherent state representing the system i.e. $a|\zeta\rangle=\zeta |\zeta\rangle$ and $|\eta_k \rangle$ be the coherent state representing the state of  bath, then we can write the SK action in Fourier space as
\begin{equation}\label{S}
	S=S_{\zeta} +S_{\eta}+S_{\zeta \eta}.
\end{equation}
$S_{\zeta}$ is the action for  system and can be written as
{\small 
\begin{eqnarray}
	\!\!\!S_{\zeta}= \!\!\!\int d\omega \begin{pmatrix}
		\zeta^*_c,\zeta^*_q 
	\end{pmatrix}
	\begin{pmatrix}
		0 & \omega -\omega_0 -{\rm i}\Gamma_M \\
		\omega -\omega_0 -{\rm i}\Gamma_M & 2{\rm i}\Gamma_M
	\end{pmatrix}
	\begin{pmatrix}
		\zeta_c \\
		\zeta_q 
	\end{pmatrix},
\end{eqnarray}}
where $\zeta_{c/q}=\zeta_{c/q}(\omega)$. For the bath, we have
{\small \begin{eqnarray}
	\!\!\!S_{\eta}= \!\!\!\sum_k\int d\omega \begin{pmatrix}
		\eta^*_{kc},\eta^*_{kq} 
	\end{pmatrix}
	\begin{pmatrix}
		0 & \omega -\omega_k -{\rm i}\epsilon \\
		\omega -\omega_k -{\rm i}\epsilon & 2{\rm i}\epsilon
	\end{pmatrix}
	\begin{pmatrix}
		\eta_{kc} \\
		\eta_{kq}
	\end{pmatrix}.  \nonumber\\
\end{eqnarray}}
Here also $\eta_{kc/q}=\eta_{kc/q}(\omega)$. The $\epsilon \rightarrow 0^+$ serves as regularization parameter. The interaction between system and bath is represented through $S_{\zeta \eta}$ and is given by
\begin{eqnarray}
	S_{\zeta \eta}&=& -\sum_k \int d\omega\Big[ g_k( \zeta^*_c(\omega) \eta_{kq}(\omega) + \zeta^*_q(\omega) \eta_{kc}(\omega) \Big. \nonumber\\ 
	&& \Big.~~~~~+ \zeta_c(\omega) \eta_{kq}(-\omega)+\zeta_q(\omega)\eta_{kc}(-\omega) ) + c.c.\Big],
\end{eqnarray}
where c.c. means complex conjugate. We now integrate out the $\eta$-fields using the Gaussian integration to get the effective action for the $\zeta$-fields i.e. for the system: $Z_{\rm eff}=\int D[\eta^*_{c,q},\eta_{c,q}]e^{iS_{\rm eff}[\eta^*_c,\eta^*_q,\eta_c,\eta_q]}$ where
\begin{eqnarray}
	S_{\rm eff}=  \int d\omega~ \zeta^\dagger_4(\omega) \begin{pmatrix}
		0 & \mathcal{D}^{\rm ad}(\omega)\\ 
		\mathcal{D}^{\rm re}(\omega) &  \mathcal{D}^K(\omega)
	\end{pmatrix} \zeta_4 (\omega).
\end{eqnarray}
Here, the four component vector $\zeta_4(\omega)=[\zeta_c(\omega), \zeta_c(-\omega), \zeta_q(\omega), \zeta(-\omega) ]^{\rm T}$, ${\rm T}$ is  transpose.  $\mathcal{D}^{\rm re},~ \mathcal{D}^{\rm ad},~ \mathcal{D}^K$ represent respectively, the inverse of  retarded, advanced and Keldysh Green's functions. These Green's functions can be explicitly written as
\begin{eqnarray} \label{keldysh component}
	\mathcal{D}^K = 2{\rm i} \Gamma_M \  {\rm diag(1,1)}	 
\end{eqnarray} 
and 
	{\small  \begin{eqnarray}{\label{eff green}}
			\mathcal{D}^{\rm re}= \begin{pmatrix}
				\omega -\omega_0 +{\rm i} \Gamma_M+ \sum (\omega) & \sum (\omega)\\
				[\sum (-\omega)]^* & -\omega -\omega_0 -{\rm i} \Gamma_M +[\sum (-\omega)]^*
			\end{pmatrix}.\nonumber\\
\end{eqnarray}}
$\Sigma(\omega)$ represents the self energy and its explicit form is given by 
\begin{eqnarray}
	\Sigma (\omega) = - \frac{1}{2} \sum_k \frac{|g_k|^2 \omega_k}{\omega^2 - {\omega_k}^2}.
\end{eqnarray}
Next, we define the bath spectral density $\mathcal{J}(\omega) =\sum_k |g_k|^2  \  \delta(\omega - \omega_k)$   so that its form can be written phenomenologically as having Lorentzian shape \cite{zubairy}
\begin{eqnarray}
	\mathcal{J}(\omega) = \frac{1}{2 \pi} \frac{\alpha \lambda^2}{(\omega - \omega_0)^2+\lambda^2}.
\end{eqnarray}
Here, $\alpha$ can be considered as the effective system-bath coupling, $\lambda$ represents the spectral width of the bath. Therefore,  $\Sigma(\omega)$ can be simplified to the form 
\begin{eqnarray}
	\label{selfenergy}
	\Sigma (\omega) &=& - \frac{1}{2}\int d\omega^{\prime} \frac{\omega^{\prime} \ \mathcal{J}(\omega^{\prime})}{\omega^2 - \omega^{\prime 2}} \nonumber\\
	&=& \frac{\alpha \lambda \omega_0}{4 }\bigg[ \frac{\omega_0^2+\lambda^2-\omega^2}{(\omega^2+\omega_0^2+\lambda^2)^2-4\omega_0^2 \omega^2}\bigg].
\end{eqnarray}
Next, we obtain  spectrum of the  system from ${\rm det}\mathcal{D}^{\rm re}=0$, and we get the following dispersion relation 
\begin{eqnarray}
	\label{char}
	\omega_{\pm}=-{\rm i}\Gamma_M \pm \sqrt{\omega_0^2-2\omega_0\Sigma(\omega)}.
\end{eqnarray}
In order to plot these solutions, we first take note of parameters involved. These parameters are energy scale of the system $\omega_0$, spectral width of the bath $\lambda$, Markovian decay rate $\Gamma_M$. The relaxation time scale for the system is given by $\tau_S \sim \omega_0^{-1}$ while the bath correlation time scale is $\tau_B\sim \lambda^{-1}$. The competition of these scales determine the Markovian or non-Markovian behavior in the system. If $\tau_S>\tau_B$, the dynamics is Markovian while vice-versa implies non-Markovian effects.  We therefore, define two new parameters $\mathcal{Q}=\omega_0/\lambda$ and $\mathcal{R}=\lambda /\Gamma_M$. Therefore, in terms of $\mathcal{Q}$ and $\mathcal{R}$, we identify $\mathcal{Q}<<1$, as Markovian regime while the non-Markovian regime is $\mathcal{Q}>>1$, Additionally, a weaker condition but not necessary, can be imposed through values of $\mathcal{R}$. For $\mathcal{R}>1$ would identify as Markovian while $\mathcal{R}<1$ as non-Markovian regime.  In terms of these parameters, we rewrite the  self energy function in the following ($z\rightarrow \omega/\lambda$):
\begin{eqnarray}
	\Sigma(z)=\frac{\alpha \mathcal{Q}}{4}\Bigg[\frac{\mathcal{Q}^2-z^2+1}{(z^2+ \mathcal{Q}^2+1)^2- 4\mathcal{Q}^2 z^2}\Bigg],
\end{eqnarray}
and the dispersion relation becomes $z=-{\rm i} \mathcal{R}^{-1}\pm \sqrt{\mathcal{Q}^2-2\mathcal{Q}\Sigma(z)}$.
In the limiting cases where $\mathcal{Q}>>1$ and $\mathcal{Q}<<1$, we have
\begin{eqnarray}
	\Sigma(z)=\begin{cases}
		\frac{\alpha \mathcal{Q}}{4(\mathcal{Q}^2-z^2)} & \mathcal{Q}>>1,\\
		\frac{\alpha \mathcal{Q}}{4}[\frac{1-z^2}{(1+z^2)^2}] & \mathcal{Q}<<1.
	\end{cases}
\end{eqnarray}
Now we see that for $\mathcal{Q}>>1$, the $\Sigma(z)$ peaks around $z=\pm \mathcal{Q}$ i.e $\omega=\pm \omega_0$. Therefore,  for a  narrow bath spectral density, characteristic equation becomes
\begin{eqnarray}
z^4+(\mathcal{R}^{-2}+2\mathcal{Q}^2)z^2-\mathcal{Q}^2(\alpha+\mathcal{R}^{-2})=0.
\end{eqnarray}
The roots of this equation are $\omega_{\pm}=\pm \sqrt{z_{\pm}}$ with $z_{\pm}=-0.5(\mathcal{R}^{-2}+2\mathcal{Q}^2)\pm0.5\sqrt{(\mathcal{R}^{-2}+2\mathcal{Q}^2)^2+ 4\mathcal{Q}^2(\alpha+\mathcal{R}^{-2})}$. Since the sign of imaginary part of the eigen frequency $\omega_{\pm}$ provides the stability of  steady state solution. A negative imaginary part of $\omega_{\pm}$ represents the stable solution while the positive part of $\omega_{\pm}$ means an unstable solution.  Therefore, there exist at least one root with positive imaginary part, hence unstable. In case of $\mathcal{Q}<<1$, the bath has no substantial effect on the system and the spectrum remains unchanged at $\omega_{\pm}\sim -{\rm i}\Gamma_M\pm \omega_0$. Furthermore, we define critical coupling $\alpha_c$ to be the point where at least one of the roots of the equation \ref{char}  vanishes i.e. $\omega=0$ and it occurs at $\alpha_c= \frac{2[\mathcal{Q}^2+1][(\mathcal{Q})^2+\mathcal{R}^2]}{\mathcal{Q}^2}$.
\begin{figure}[t] 
	\includegraphics[width=2.7cm,height=3cm]{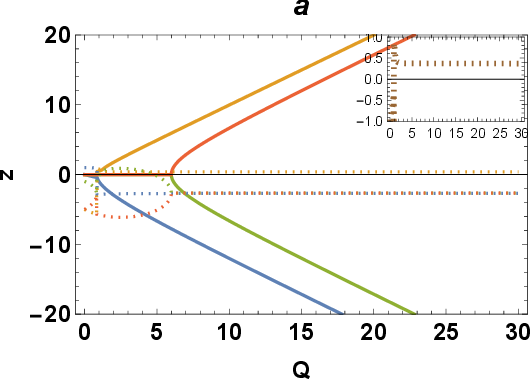}\hspace{0.1cm}\includegraphics[width=2.7cm,height=3cm]{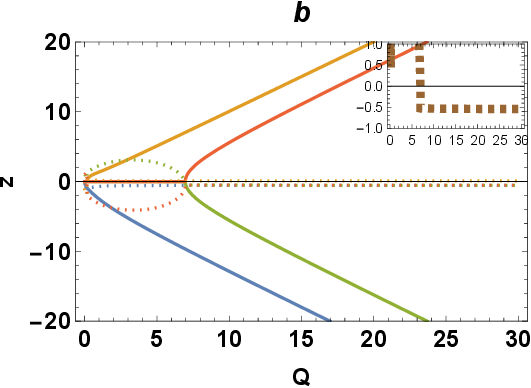}\hspace{0.1cm}\includegraphics[width=2.7cm,height=3cm]{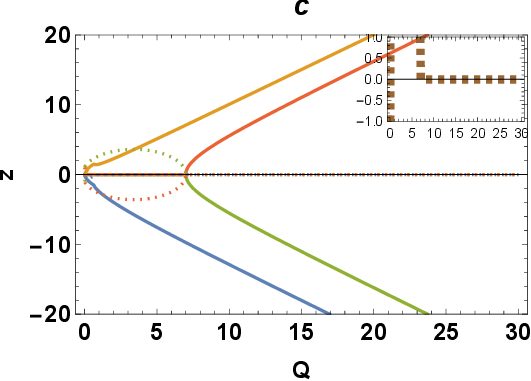}
	\caption{ We plot  real and imaginary parts of  roots of the characteristic equation \ref{char} with respect to $\mathcal{Q}\equiv \omega_0/\lambda$ for different values of $\mathcal{R}\equiv \lambda/\Gamma_M$:(a) $\mathcal{R}=0.1$ (b) $\mathcal{R}=1$ and (c) $\mathcal{R}=10$.Dotted lines are the imaginary part while the solid lines represent the real part of the roots. Inset in each figure represents the plot of imaginary part of the unstable root, and this root changes its sign from positive value for $\mathcal{R}=0.01$ (unstable) to negative for $\mathcal{R}=10$(stable). }
	\label{Fig1}
\end{figure}
In fig \ref{Fig1}, we plot real and imaginary parts of the roots of equation (\ref{char}) with respect to $\mathcal{Q}$ at large $\alpha $ values and $\mathcal{R}=0.1,~1,~10$. The imaginary parts of the roots are shown by dotted lines while the solid lines correspond to real part of the roots. We observe that there exist at least one root with positive imaginary part (as shown in the inset of  figure \ref{Fig1}) signaling an instability in the system. This instability is reflected in the divergence of  number density. The system can be stabilized via tuning the values of $\mathcal{R}$. As we progress from {$\mathcal{R}=0.1$} to $\mathcal{R}=10$, the imaginary part of the unstable root changes sign from positive to negative as shown in the inset of figure \ref{Fig1}(a)-(c). Therefore, in the non-rotating wave approximation, the condition $\mathcal{R}<1$ (non-Markovian) is unstable towards a transition.

\begin{figure}[t] 
	\includegraphics[width=3.0in,height=2.0in]{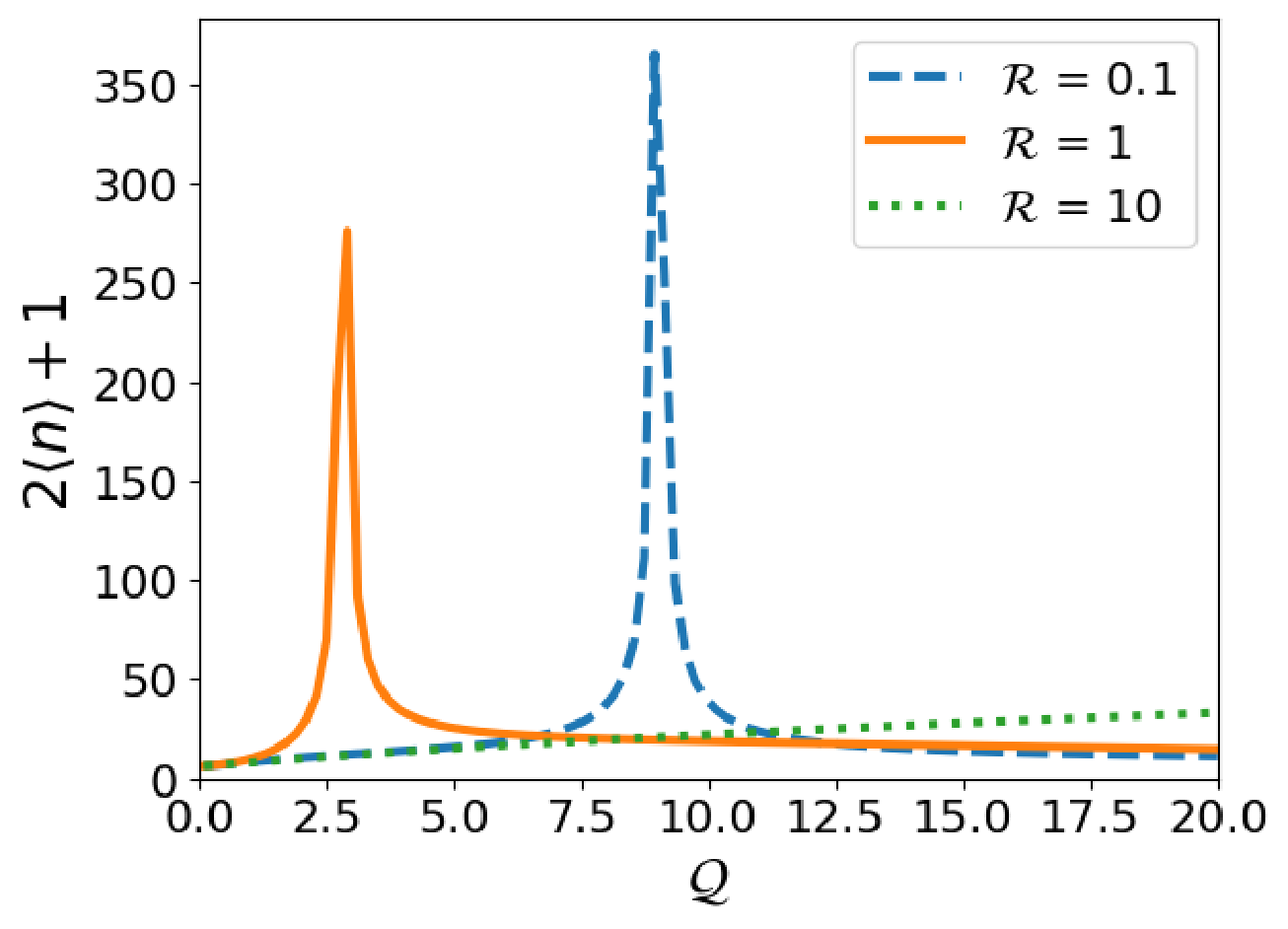}
	\caption{In this figure, we plot $2\langle n\rangle+1$ with respect to $\mathcal{Q}$ for different values of $\mathcal{R}$. As we progress through $\mathcal{R}<1$ to $\mathcal{R}>>1$, the transition point changes with no transition in steady state for $\mathcal{R}>1$.    }
	\label{Fig2}
\end{figure}

\subsection{Correlations}\label{Correlations}
Next, we evaluate the steady state one point  correlation function that yields the number density. The one point correlation function is related to Keldysh Greens function in the following way:
\begin{eqnarray}
	C (t,t^{\prime})= \langle \{a(t),a^{\dagger}(t^{\prime})\}\rangle ={\rm i}G^K(t,t^{\prime}).
\end{eqnarray}
where $\{A,B\}=AB+BA$ defines anti-commutator. The Keldysh Greens function is obtained from $\mathcal{D}^K$ using the formula $G^K= -G^{\rm re}\mathcal{D}^K G^{\rm ad}$, {with $G^{\rm re}(\omega)= [\mathcal{D}^{\rm re}(\omega)]^{-1}$ and $G^{\rm ad}(\omega)=[G^{\rm re}(\omega)]^*$. In the steady state, the system becomes time translation invariant and therefore correlation function depends only on time differences $t-t'$, which implies for equal times, the correlation function $C(t,t')=C(0)=\langle \{a,a^{\dagger}\}\rangle=2\langle n\rangle+1$;  $\langle n\rangle =\langle a^{\dagger}a\rangle$ is the  average number density}. Therefore, in the steady state, we compute the distribution function as follows:
\begin{eqnarray}
	2\langle n\rangle  +1 = {\rm i}\int \frac{d\omega}{2\pi} G^K_{11}(\omega),
\end{eqnarray}
where the first diagonal entry of the Keldysh Greens function is given by 
\begin{eqnarray}
	{\rm i}G^K_{11} (\omega) = \frac{2\Gamma_M[(\omega+ \omega_0 -\Sigma)^2 + \Gamma_M^2 + \Sigma^2 ]}{(\omega^2 -\Gamma_M^2 -\omega_0^2 + 2 \omega_0 \Sigma)^2 + 4\omega^2 \Gamma_M^2}.
\end{eqnarray}
In figure \ref{Fig2}, we have plotted $2\langle n\rangle +1$ with respect to $\mathcal{Q}$ for different $\mathcal{R}$ values. {The divergence in  number density reflects the macroscopic occupation of the bosonic mode at different $\mathcal{Q}$ values implying an instability in the system. Since the present model is the effective model obtained from the  thermodynamic limit ($N\rightarrow\infty$) of Dicke model, this transition is well defined and system enters into  superradiant phase with $\langle a\rangle \ne 0$ \cite{SPec,pt,PK}. However, if the system is simply a single oscillator, this divergence reflects dynamic instability (not a phase transition) in the system. This instability can be attributed to the resonance effect at strong system-bath coupling \cite{Agarwal_2012}. Under the strong non-Markovian effects, the backflow of information leads to an uncontrolled energy dissipation causing the number density to diverge}. This transition is pronounced at {$\mathcal{R}=0.1<1$}-value for large $\mathcal{Q}$-values. As we increase $\mathcal{R}$-values the transition point shifts to lower $\mathcal{Q}$ and finally for $\mathcal{R}>>1$, the transition disappears. This result is consist from the dispersion equation \ref{char} (figure \ref{Fig1}(a)-(c)). In the region $\mathcal{R}>1$, the system becomes stable for $\mathcal{Q}>1$ values and thus no region of transition. Thus we conclude that the system can be tuned to phase transition via tuning system from Markovian to non-Markovian.

Next, we look at the fluctuation-dissipation relationship  in the steady state. This is reflected through the equation \cite{sieberer2}
\begin{eqnarray}
	G^K(\omega) ={ G^{\rm re}(\omega)} \circ \mathcal{F}(\omega) - \mathcal{F}(\omega) \circ {G^{\rm ad}(\omega)},
	\label{FDT}
\end{eqnarray}
where $\mathcal{F}(\omega)$ is called as distribution function that has the form of $2{\langle n \rangle}+1=\coth\frac{ \beta \omega}{2}$ in equilibrium. The eigen values of $\mathcal{F}(\omega)$ provide the quantitative description of the effective temperature. Solving this equation \ref{FDT}, we get $\mathcal{F}(\omega)= \sigma_z+\frac{1}{\omega}\Sigma(\omega) \sigma_x$. The effective temperature is calculated as the dimensional coefficient of $1/\omega$ in the expansion of $\mathcal{F}(\omega)$ in powers of $1/\omega$. Thus, {we have an }effective temperature  $T_{\rm eff}= \alpha$ independent of $\mathcal{Q}$ and $\mathcal{R}$.

\section{Quantum Zeno Effect}\label{QZE}

The phenomena of quantum Zeno and anti-Zeno effects provide an enlightening framework linking measurement with environmental couplings and back-action dynamics due to such dissipative processes \cite{itano,ballentine}. QZ effect describes freezing out the dynamics while AQZ effect  represents the enhancement of dynamic evolution induced by frequent measurements\cite{fischer}. Different experiments realized  QZ effects, for example systems like trapped ions\cite{itano}, superconducting qubits \cite{harrington}, Ultracold atoms\cite{fromel}, semiconductor systems. QZ effects  has been used to extend the lifetime of molecular states, improve the spectroscopic precision \cite{kofman} etc. It has also been proposed that the Zeno effect could be used to preserve the entanglement between two atoms\cite{maniscalco}. The goal of these studies has been to comprehend how measurements affect quantum system dynamics and how this knowledge may be used to manipulate and control quantum states.

In the steady state, we can quantify Zeno effect through Zeno parameter defined as follows \cite{thapliyal,thapliyal2}:
\begin{eqnarray}
	\xi =\frac{ \langle n\rangle - \langle n\rangle |_{\alpha=0}}{\langle n\rangle },
\end{eqnarray}
where $\left\langle n\right\rangle$ is the steady state number density. In Keldysh formalism, Zeno parameter {$\xi$} can be written as:
\begin{eqnarray}
	\label{ZENO}
	\xi &=&\frac{[2\langle n \rangle + 1]-[2\langle n \rangle|_{\alpha=0} + 1]}{[2\langle n \rangle + 1]-1}, \nonumber\\
	&=& \frac{ {\rm i}\int d\omega \Big[G_{11}^K(\omega,\alpha)- G_{11}^K(\omega,\alpha=0)\Big]   }{{\rm i}\int d\omega G_{11}^K(\omega,\alpha)-1}.
\end{eqnarray}
The values of $\xi$ determine whether the dynamics is Zeno or anti-Zeno in nature.
A positive (negative) value of $\xi$ indicates the anti-Zeno(Zeno) effect \cite{naikoo} i.e. there is  enhancement(suppression) in the average photon numbers  associated with the bosonic mode due to its coupling $\alpha$ with the bath. 
\begin{figure}[t] 
	\includegraphics[width=4.0cm,height=4.0cm]{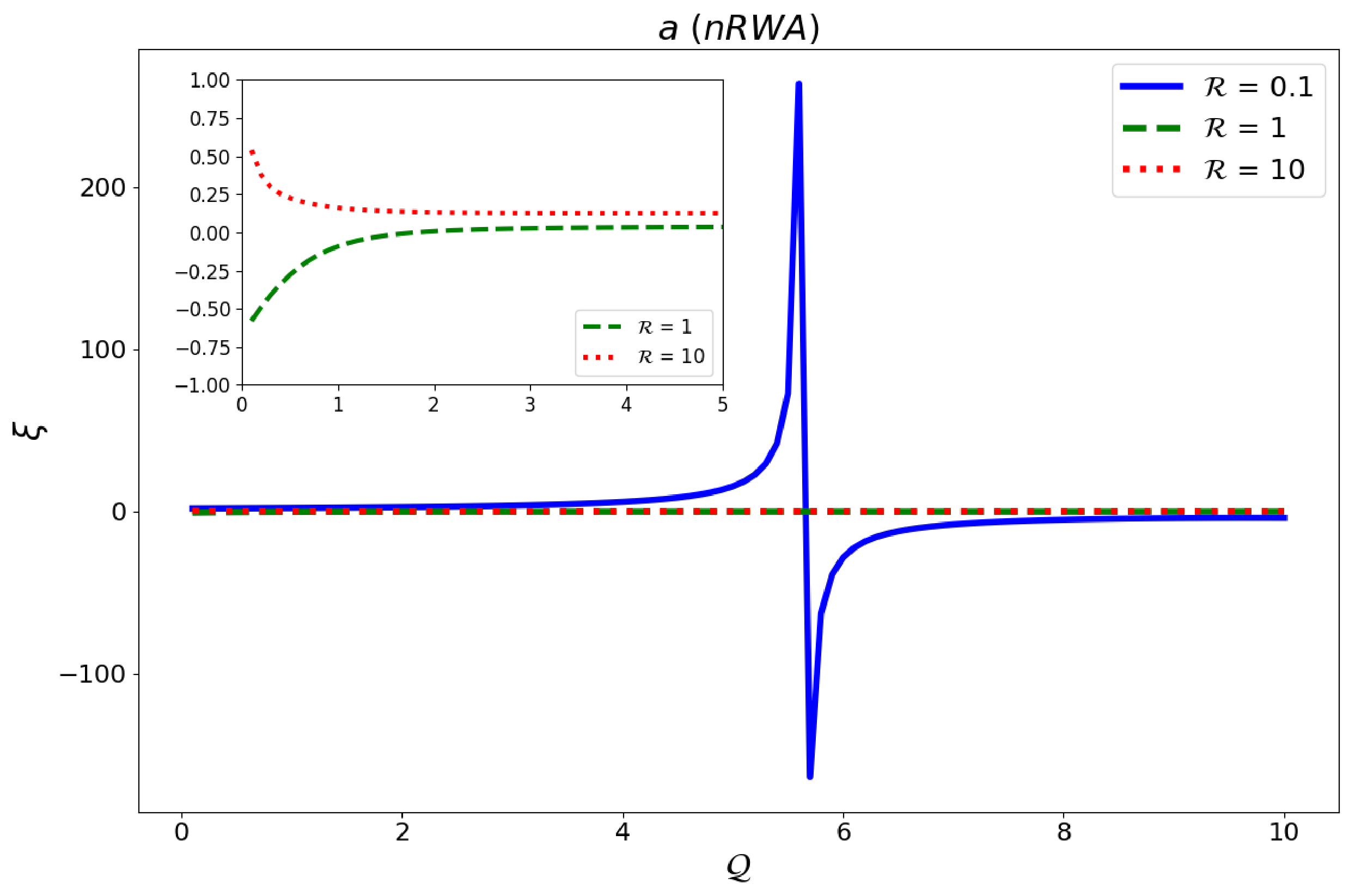}\hspace{0.2cm}
	\includegraphics[width=4.0cm,height=4.0cm]{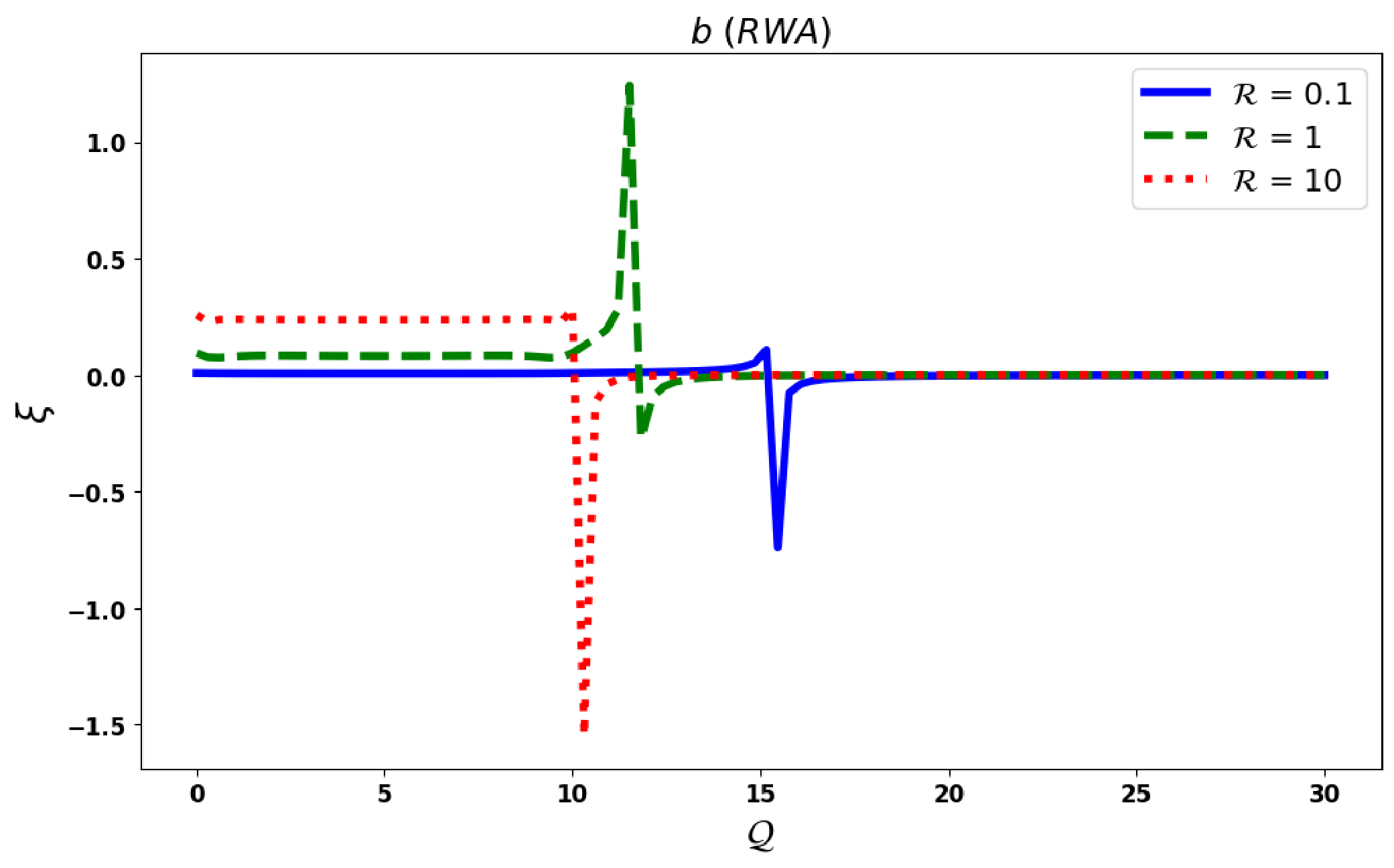}
	\caption{These plots depict the steady state Zeno parameter $\xi$ in (a) non-rotating wave approximation (nRWA) (b) rotating wave approximation (RWA). We plot Zeno parameter $\xi$ with respect to $\mathcal{Q}$ for $\mathcal{R}=0.1,~1,~ 10$.
 (a) We can see for $\mathcal{R}=0.01$  the system enters into the anti-Zeno region ($\xi>0$) first then making transition to Zeno region with $\xi<0$. In case of nRWA, we observe that (shown in inset of (a)) for  $\mathcal{R}=1$ values system starts in Zeno region while for large $\mathcal{R}=10$, we have anti-Zeno effect that finally vanishes as $\mathcal{Q}$ increases.  (b) In  RWA case, $\mathcal{R}>1$, the system starts initially in the anti-Zeno region and finally making transition to Zeno region. However, {$\mathcal{R}=0.1$}, the system  show only Zeno effect.}
	\label{Fig3}
\end{figure}

In figure \ref{Fig3}(a), we plot steady state Zeno parameter $\xi$ (non-rotating wave approximation), with respect to $\mathcal{Q}$ for different values of $\mathcal{R}$. From this figure we observe that for small {$\mathcal{R}=0.1$}, dynamics of the system enters first into anti-Zeno region ($\xi>0$)  then followed by a transition to Zeno effect ($\xi<0$). As we further increase $\mathcal{Q}$-values, the system does not evolve to particular dynamics i.e. $\xi=0$. The inset of  figure \ref{Fig3}(a) represents the variation of $\xi$ with $\mathcal{Q}$ for  $\mathcal{R}=1,10$. We observe in case of $\mathcal{R}=1$, system has initially Zeno dynamics ($\xi<0$) while for $\mathcal{R}=10$, the system posses anti-Zeno ($\xi >0$) dynamics, and as we tune $\mathcal{Q}$ to large values, these effects dies out without showing any transition as shown for the case of {$\mathcal{R}=0.1$. This behaviour can be attributed to the collective role played by the non-Markovianity induced by the bath; and competition between dissipative and coherent dynamics in the system. These effects are characterized through the parameters $\mathcal{Q}=\omega_0/\lambda$ and $\mathcal{R}=\lambda/\Gamma_M$. 
	The parameter $\mathcal{R}$ in case of fixed dissipation rate $\Gamma_M$, its large values reflects the broader spectral width $\lambda-$ a Markovian nature. Therefore, as the system enters into the Markovian dynamics, the backflow of information weakens leading to the slower dynamics. This implies, within Markovian regime system {does not show any QZ} or AQZ effect.  While in non-Markovian case $\mathcal{Q}>1,\mathcal{R}<1$, there is a continuous backflow of information to the system leading to the transitions between AQZ and QZ regimes.
    }

\section{ Rotating Wave Approximation}\label{RWA}

In this section, we consider the RWA Hamiltonian given by
\begin{eqnarray}
	H=H_A+ H_B + \sum_k(g_kab_k^{\dagger}+ g_k^* a^{\dagger} b_k),
\end{eqnarray}
and compare the results with that of non-RWA. After tracing out the bath degrees of freedom as done earlier, we arrive at the  diagonal inverse retarded Greens function $\mathcal{D}^{\rm re}_{\rm RWA}$: 
{\small	\begin{eqnarray}{\label{eff green}}
		\mathcal{D}^{\rm re}_{\rm RWA}= \begin{pmatrix}
			\omega -\omega_0 +{\rm i} \Gamma_M+ \tilde{\Sigma}(\omega) & 0\\
			0 & -\omega -\omega_0 -{\rm i} \Gamma_M +[\tilde{\Sigma}(-\omega)]^*
					\end{pmatrix}.\nonumber \\
\end{eqnarray}}
In RWA, the self energy $\tilde{\Sigma}(\omega)$ is given by 
\begin{equation}
	\tilde{\Sigma}(\omega) = - \frac{1}{2} \sum_k \frac{|g_k|^2}{\omega-\omega_k}.
\end{equation}
In terms of parameter $\mathcal{Q}$ and $\mathcal{R}$, this self energy can be simplified to the form
\begin{eqnarray}
	\tilde{\Sigma}(z)= \frac{\alpha }{2} \frac{\mathcal{Q}-z}{(\mathcal{Q}-z)^2+1}.
\end{eqnarray}
The dispersion relations in term of $\mathcal{Q}$ and $\mathcal{R}$ is given by \begin{eqnarray}
	\label{DRWA}
	z+{\rm i}\mathcal{R}^{-1} -\mathcal{Q}+ \tilde{\Sigma}(z)&=&0,\nonumber\\
	z+{\rm i}\mathcal{R}^{-1} +\mathcal{Q}- [\tilde{\Sigma}(-z)]^*&=&0.
\end{eqnarray}
\begin{figure}[t] 
	\includegraphics[width=2.7cm,height=3cm]{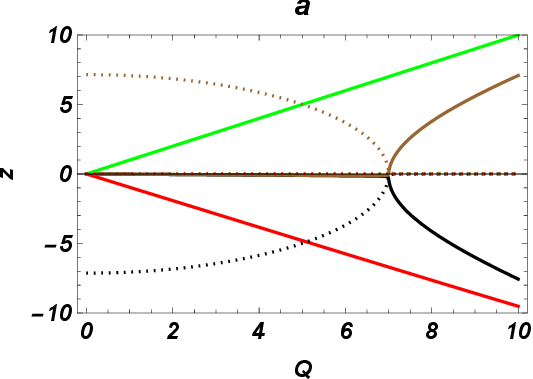}\hspace{0.1cm}\includegraphics[width=2.7cm,height=3cm]{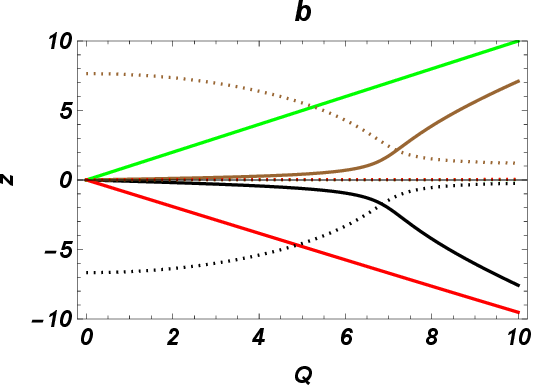}
	\hspace{0.1cm}\includegraphics[width=2.7cm,height=3cm]{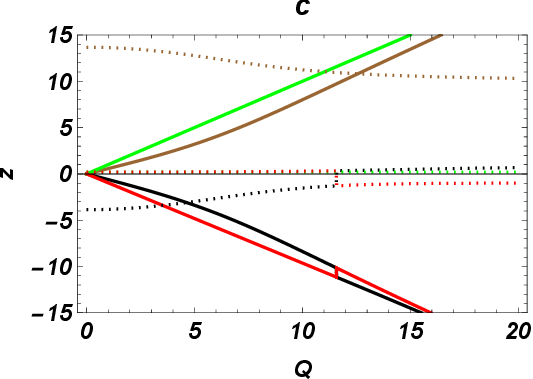}
	\caption{ We plot  real and imaginary parts of  roots of the characteristic equation \ref{char} with respect to $\mathcal{Q}\equiv \omega_0/\lambda$ for different values of $\mathcal{R}\equiv \lambda/\Gamma_M$:(a) $\mathcal{R}=0.1$ (b) $\mathcal{R}=1$ and (c) $\mathcal{R}=10$. Dotted lines are the imaginary part while the solid lines represent the real part of the roots. }
	\label{FigRWA}
\end{figure}
In figure \ref{FigRWA}(a)-(c), we plot the roots of the dispersion relation equations \ref{DRWA} as a function of $\mathcal{Q}$ for {$\mathcal{R}=0.1,1,10$}. We observe that there exist at least one root with positive imaginary part making it unstable. Since $\mathcal{R}<1$ identifies as non-Markovian region, from figure \ref{FigRWA}(a), the positive imaginary part vanishes at a value  $\mathcal{Q}>1$
and at this value of $\mathcal{Q}$ its real part becomes non-zero (positive). As we move away from $\mathcal{R}<1$ to $\mathcal{R}>1$, figures \ref{FigRWA}(b)-\ref{FigRWA}(c), we see that the same root does not vanish for very large values of $\mathcal{Q}$, thus making the system unstable.Therefore, in rotating wave approximation, non-Markovian regime $\mathcal{Q}>1,\mathcal{R}<1$ describes the  stable state of the system while Markovian dynamics implies instability in the system.  

This instability would imply a divergence in the number density reflecting the phase transition in the system. However, if we calculate effective temperature of the system using same procedure as above we get $\mathcal{F}=\sigma_z$, $\sigma_z$ is pauli spin matrix. This implies, we have  $T_{\rm eff}=0$. Thus in comparison to non rotating wave approximation where the steady state has non-zero local effective temperature, the system in case of RWA thermalizes at $T_{\rm eff}=0$.

\textit{Zeno Effect}: Since rotating wave approximation Hamiltonian has different symmetry in comparison to non-RWA counter part. Therefore, we look at the behaviour of the Zeno effect in RWA under different parameter values. We now plot in figure \ref{Fig3}(b) Zeno parameter $\xi$ as given in the equation \ref{ZENO} in the rotating wave approximation as a function of quality factor $\mathcal{Q}$ for different values of $\mathcal{R}$. We observe in the non-Markovian regime $\mathcal{Q}>1,\mathcal{R}<1$, the system shows the quantum Zeno effect for a particular value of $\mathcal{Q}$, while in the regions with $\mathcal{R}=1,~10$, the system starts the dynamics in the anti-Zeno regime with a transition to the Zeno regime. Compared to non-RWA, where the zeno and anti-zeno effect were absent in the Markovian regime, RWA exhibits an anti-zeno effect in the Markovian regime. The transition from the {AQZ} to QZ regimes is observed in the RWA for all considered values of $\mathcal{R}$.

\section{Conclusions}
In conclusion, we have studied the damped harmonic oscillator represented by a decaying bosonic mode coupled to a non-Markovian bath described by Lorentzian spectral density. Using the Schwinger-Keldysh formalism, an effective action for the system is obtained after integrating out the bath degrees of freedom. The steady state properties of the system were studied in rotating as well in non-rotating wave approximations. We characterized the dynamics through two parameters $\mathcal{Q}=\omega_0/\lambda$ (quality factor) and $\mathcal{R}=\lambda/\Gamma_M$. Using these parameters, we identified $\mathcal{Q}>1,\mathcal{R}<1$ as non-Markovian while the $\mathcal{Q}<1,\mathcal{R}>1$ as Markovian regimes of the dynamics. A detailed study of spectrum of the system in non-RWA and RWA revealed that the system undergoes an instability towards a phase transition  in former case for non-Markovian region while in the later case (RWA), the Markovian region is unstable.
Therefore, tuning the quality factor $\mathcal{Q}$, the system can be stabilized under both dynamics. Furthermore, we have shown that the system in non-RWA thermalizes at non zero effective temperature set by intrinsic system-bath coupling $\alpha$ independent of the $\mathcal{Q}$ and $\mathcal{R}$ while in case of RWA, this effective temperature vanishes.

Next, we studied the steady state quantum Zeno and anti-Zeno effects characterized by Zeno parameter $\xi$. We have shown that in case of non-RWA, for  $\mathcal{R}>1$ values the system dynamics is initially in Zeno regime while for $\mathcal{R}=1$ values, it is in ant-Zeno regime. As we increase quality factor $\mathcal{Q}$ for these $\mathcal{R}$-values, the system shows no Zeno or anti-Zeno effects.  Also for complete non-Markovian case $\mathcal{Q}>1,\mathcal{R}<1$, the system enters into anti-Zeno at particular value of $\mathcal{Q}$ then making a transition to Zeno region. However, in case of RWA, we observed that the dynamics is initially anti-Zeno for $\mathcal{R}\ge 1$ and finally making a transition to Zeno region. While the purely non-Markovian dynamics in case of RWA has no appreciable anti-Zeno effect. Thus, the non-Markovinity parameters $\mathcal{Q}, \mathcal{R}$ play an important role in tuning the system dynamics \cite{mqlac}.

 \bibliographystyle{apa}
\bibliography{refer}	
\end{document}